\documentclass[twocolumn,pra,floatfix]{revtex4}
\usepackage{amssymb}
\usepackage{soul}
\usepackage{graphicx}

\usepackage{color}
\usepackage{bm}

\newcommand{\hc}{\mathrm{H.c.}}
\newcommand{\trf}{\mathrm{tr}_F}

\begin{document}

\title{Dynamics of a two-level atom in the presence of a medium-assisted thermal field}
\author{Razieh Gonouiezadeh and Hassan Safari}

\affiliation{Department of Photonics, Graduate University of Advanced Technology, Kerman, Iran}

\begin{abstract}
In this paper the time evolution of a two-level atom in the presence of medium-assisted thermal
field is explored through which, the formula of decay rate of an excited atom is generalized 
in two aspects. The obtained formula applies for the thermal electromagnetic field as well as the 
presence of arbitrary arrangement of magneto-electric media. In order to be general with respect to the material environment, the Green's function approach is used. It is seen that the non-zero temperature contributes to the decay rate via an 
additive term that is equal to the zero-temperature result multiplied by two times of photon number
at atomic transition frequency.

\end{abstract}

\maketitle

\section{Introduction}
\label{intro}
\noindent
 The decay rate of an excited atom, $\Gamma$, 
identifying the probability rate of its transition to a lower state, 
stands as a fundamental concept in atomic physics. The lifetime of atomic 
excitation  $\tau$, being defined as $1/\Gamma$, is a measure of the 
intrinsic instability of specific atomic states. 
However, its stemming from the interaction of the atom with the 
fluctuating quantized 
electromagnetic field around, somehow 
differs from the ordinary sense of being intrinsic \cite{berestetskii1982quantum}.
It was first anticipated by Einstein in a series 
of papers  using Planck's concept of 
black-body radiation spectrum \cite{einstein1996}. 
The pioneering dynamical analysis was the one 
introduced by Weisskopf and Wigner \cite{Weisskopf1930} via an 
approach based upon Schr\"{o}dinger 
equation, leading to the excitation 
decay rate. Among the other approaches used to the theory are 
Fermi's golden rule \cite{Barnett1992, loudon2000quantum}, 
density operator approach \cite{Scully1999quantum}, 
and informational-entropic approach for hydrogen-like atoms 
\cite{gleiser2018predicting}. 

In the original studies the atom is considered in an infinitely extended 
free space while the electromagnetic field is in its ground state (vacuum state). Because of infinite number of degrees of freedom for 
the electromagnetic field in general, there are many ways in which the 
emitted photon can be found in result of a single atomic transition and 
hence, the process is irreversible. The atom-filed interaction was developed to 
field mode in optical cavity by Jaynes and Cummings 
\cite{jaynes1963comparison} leading to the non-intuitive prediction of 
that the decay rate may be altered by the presence of material boundaries, 
because of the change in the density of the field modes. 

Later on, the theory was developed in 
many aspects, among them, extension to atoms embedded in infinitely extended 
homogeneous media \cite{nienhuis1976atomic}, in the neighbourhood of a 
dielectric surface \cite{khosravi1991vacuum}, in
the presence of arbitrarily arranged dispersive and dissipative dielectric environment \cite{barnett1996decay, scheel1999spontaneous, dung2000spontaneous}, accounting for the local-field effects
\cite{scheel1999quantum},
modification to
magneto-electric media \cite{dung2003electromagnetic}
 and applying the method to spherical cavity and planar half-space \cite{dung2000spontaneous}, and
atoms in causal plasma \cite{pishbin2012diss, pishbin2012mag}. 

The other scenarios studied are atom in the presence 
of conducting wedge \cite{mohammadi2015energy}, in nano-bubbles \cite{klimov1999enhancement}, in two-dimensional micro-cavities \cite{zhang2012temperature}, atoms moving in a relativistic velocity 
\cite{papadatos2020relativistic}, extension to multi-level 
atoms \cite{Zhu1995, buhmann2004nonper}, and to 
thermal or squeezed fields 
\cite{gardiner1986inhibition, cirac1991two, garraway1997decay, Scully1999quantum, zhang2012temp}. Thermalisation and spontaneous decay of a two-level atom beyond the Markovian approximation has also been studied \cite{salo2006non}. The atomic excitation decay rate
as a direct consequence of field fluctuation, has been suggested as
a measure for sub-vacuum phenomena of non-classical field state \cite{ford2011effects}.

The study of decay rate of an atom has been of interest also in experimental side.
It was shown that the decay rate is significantly affected by the surrounding media \cite{rikken1995enhancement, snoeks1995measuring, yablonovitch1988inhibited}, and the observation of local field effects were emphasised for the case of atom being embedded in medium \cite{rikken1995local, schuurmans1998local}.

Applications of such research in lasers, quantum information and sensors can be highlighted as notable instances. For example, in quantum computers, two-level systems near linear magnetoelectric materials can be used as qubits. The transition rate in these systems plays a key role in determining the accuracy and speed of quantum operations \cite{bouwmeester2000physics}. Furthermore the real systems are unavoidably coupled with the environment in which they are embedded, hence they have to be considered as open systems, no matter how weak is the interaction. To this end, many research has been conducted.  By way of illustration a research has presented non-Markovian qubit dynamics in a thermal ﬁeld bath \cite{shresta2005non} or another one  have highlighted the importance of thermal noises on enhancing qubit quantum information \cite{gillard2018enhancing} and more recently, control of a qubit under Markovian and non-Markovian noise has been investigated \cite{delben2023control}. 

In the rapidly evolving landscape of quantum information, the necessity for such research has never been more apparent. This paper seeks to delve into dynamics of a two-level atom in the presence of a medium-assisted thermal field, aiming to shed light on some aspects of quantum information.

In this paper we extend  the previous investigations using density  operator approach, that gathers the 
previous studies that either overlooked the influence of material 
environment \cite{dung2003electromagnetic} or the impact of thermal 
electromagnetic field on the decay rate \cite{Scully1999quantum}. The resulting formula
is general with respect to the magneto-electric and geometrical properties of  (linearly responding)
surrounding media as these are to be included in the Green's tensor.

The rest of the article is organised as follows: Section \ref{formula} is devoted to introduction of the system under consideration and the quantization scheme used. In Sec.~\ref{decay} the time evolution of the 
atom interacting with a thermal bath is investigated from which, the 
decay rate of an excited atom is obtained as a prompt result. The summary 
and concluding results is given in Sec.~\ref{summ}.

\section{Basic formula}
\label{formula}

The system under consideration is comprised of an atom and a medium-assisted
electromagnetic field, for which the Hamiltonian in its multi-polar coupling form may be written as 
\begin{center}
\begin{equation}
\label{H} 
\hat{H}=\hat{H}_A + \hat{H}_F + \hat{H}_I,
\end{equation}
\end{center}
where $\hat{H}_A$, $\hat{H}_F$ and $\hat{H}_I$, indicate atomic Hamiltonian, field Hamiltonian and atom-field interaction Hamiltonian, respectively.

The atomic Hamiltonian,
with the unperturbed eigenvalues $E_i$ and corresponding eigenstates $|i\rangle$ in hand, reads \begin{equation}\label{AH} 
\hat{H}_A= \sum_{i} E_i |i\rangle \langle i|.
\end{equation}
The field Hamiltonian can be written as
\begin{equation}\label{FH} 
\hat{H}_F = \sum_{\lambda=e,m}^{ } \int d^3 r \int_{0}^{\infty} d\omega ~\hbar \omega ~\hat{\bm{f}}_{\lambda}^{\dagger}(\bm{r} , \omega ) \cdot \hat{\bm{f}}_\lambda(\bm{r} , \omega ),
\end{equation} 
where the fundamental bosonic operators $\hat{\bm{f}}_\lambda$ and $\hat{\bm{f}}_{\lambda}^{\dagger}$ are, respectively, the photon annihilation and creation opertors obeying the commutation relations 
\begin{equation}\label{com1} 
\big[\hat{\bm{f}}_\lambda(\bm{r} , \omega ),\hat{\bm{f}}_\lambda(\bm{r}^\prime , \omega^\prime )\big]=\big[\hat{\bm{f}}_{\lambda}^{\dagger}(\bm{r} , \omega ),\hat{\bm{f}}_{\lambda^\prime}^{\dagger}(\bm{r}^\prime , \omega^\prime )\big]=\bm{0},
\end{equation}
\begin{equation}
\label{com2} 
\big[\hat{\bm{f}}_\lambda(\bm{r} , \omega ),\hat{\bm{f}}_{\lambda^\prime}^{\dagger}(\bm{r}^\prime , \omega^\prime )\big]  = \delta_{\lambda\lambda^\prime} \bm{\delta}(\bm{r}-\bm{r}^\prime) \delta(\omega-\omega^\prime)
\end{equation}
with label $\lambda$ referring to the electric ($\lambda$ $\!=$ $\!e$) or magnetic ($\lambda$ $\!=$ $\!m$) nature of excitation sources of electromagnetic field \cite{dung2003electromagnetic, buhmann2004nonper}.
The third term in the right hand side of Eq.~(\ref{H}), the interaction Hamiltonian, referring to the position of the atom by $\bm{r}_A$, can be given as
\begin{equation}\label{IH} 
\hat{H}_I(t) = -\hat{\bm d} \cdot \hat{\bm E}(\bm{r}_A,t)
\end{equation}
with $\hat{\bm d}$ and $\hat{\bm E}$ being, respectively, the electric dipole moment of the atom and the electric field. 

In order to be general with respect to magnetoelectric as well as geometric properties of the material environment, we use the Green's function approach in which the frequency components of the electric 
field, $\hat{\bm{E}}(\bm{r}, \omega)$,  is given as a linear combination of operators $\hat{\bm{f}}_\lambda$ as
\begin{equation}\label{EF} 
\hat{\bm{E}}(\bm{r}, \omega) = \sum_{\lambda=e,m} \int d^3 r\,  \bm{G}_\lambda  (\bm{r}, \bm{r} , \omega) \cdot \hat{\bm{f}}_\lambda(\bm{r} , \omega ),
\end{equation}
using which
\begin{equation}\label{EF2} 
\hat{\bm{E}}(\bm{r},t) = \int_0^\infty d\omega \hat{\bm{E}}(\bm{r}, \omega)e^{-i\omega t} + \hc\,.
\end{equation}
The mode tensors $\bm{G}_e$ and $\bm{G}_m$ in Eq.~(\ref{EF}) are given in terms of the Green 
tensor $\bm{G}$ as follows
\begin{equation}\label{Ge} 
\bm{G}_e  (\bm{r}, \bm{r}^\prime , \omega)= i \frac{\omega^2}{c^2} ~\sqrt{{\frac{\hbar}{\pi \epsilon_0}} \mathrm{Im} \varepsilon(\bm{r}^\prime , \omega) } ~\bm{G} (\bm{r}, \bm{r}^\prime , \omega),
\end{equation} 
\begin{equation}\label{Gm} 
\bm{G}_m  (\bm{r}, \bm{r}^\prime , \omega)= i ~\frac{\omega}{c} ~\sqrt{{\frac{\hbar}{\pi \epsilon_0}}   \frac{\mathrm{Im}\mu(\bm{r}^\prime , \omega)}{|\mu(\bm{r}^\prime , \omega)|^2} } ~\big[\nabla' \times \bm{G}(\bm{r}', \bm{r} , \omega)\big]^\top,
\end{equation} 
where $\varepsilon$ and  $\mu$ are the relative electric permittivity and magnetic permeability of the media, respectively. The Green tensor contains all geometric and magnetoelectric properties of the environment media (via $\varepsilon$ and $\mu$) and is the unique solution to the Helmholtz equation \cite{buhmann2004nonper}
\begin{eqnarray}
\label{HE} 
&&\nabla \times \frac{1}{\mu(\bm{r} , \omega)}\nabla \times \bm{G}(\bm{r}, \bm{r}^\prime , \omega) -  \frac{\omega^2}{c^2}~\varepsilon(\bm{r} , \omega)~\bm{G}(\bm{r}, \bm{r}^\prime , \omega)\nonumber\\
&&= \bm{\delta}(\bm{r}-\bm{r}^\prime)
\end{eqnarray}
with the boundary condition
\begin{equation}\label{BC} 
\bm{G}
(\bm{r}, \bm{r}' , \omega)\rightarrow 0 \quad \textrm{for}\,
  |\bm{r}-\bm{r}^\prime|\rightarrow \infty\,.
\end{equation} 
Similar to $\varepsilon(\bm{r} , \omega)$ and  $\mu(\bm{r} , \omega)$, the Green tensor as a response function is analytic in the upper half of the complex frequency plane and obeys the Schwartz reflection principle,
\begin{equation}\label{SRP}
\bm{G}(\bm{r}, \bm{r}^\prime , \omega)=\bm{G}^\ast(\bm{r}, \bm{r}^\prime , -\omega^\ast),
\end{equation}
Onsager-Lorentz reciprocity,
\begin{equation}\label{OLR} 
\bm{G}( \bm{r}^\prime, \bm{r},\omega)=\bm{G}^\top(\bm{r}, \bm{r}^\prime , \omega),
\end{equation} 
and the integral relation
\begin{eqnarray}
\label{IE} 
&&\sum_{\lambda=e,m} \int d^3  s\, \bm{G}_\lambda(\bm{r}, \bm{s} , \omega) \cdot 
{\bm{G}_\lambda^{\ast\top}}(\bm{r}^\prime, \bm{s} , \omega)
\nonumber\\
&&\hspace*{.3in}=\frac{\hbar \mu_0}{\pi} \omega^2 \mathrm{Im} \bm{G} (\bm{r}, \bm{r}^\prime , \omega).
\end{eqnarray}

\section{Atomic dynamics}
\label{decay}
In this section we investigate the time evolution of a system consisted of an atom and an interacting electromagnetic field of temperature $T$, making use of density operator approach. To avoid unnecessary complication we restrict our calculations to the case of a two-level atom as the generalization to many-level atoms is straightforward. Referring to the ground state and excited state of the atom, respectively by $|g\rangle$ and $|e\rangle$, the electric-dipole operator of the atom reads in terms of Pauli operators
$\hat{\sigma} =|g\rangle\langle e|$ and $\hat{\sigma}^\dagger =|e\rangle\langle g|$ as
\begin{equation}\label{di}
\hat{\bm d}(t)=\bm d ~(\hat{\sigma} e^{-i\omega_{A}t}+ \hat{\sigma}^\dagger e^{+i\omega_{A}t}),
\end{equation}
where $\omega_A = (E_e - E_g)/\hbar$ is the atomic transition frequency.

Denoting by $\hat{\rho}(t)$ the density operator of the system, its time evolution is obtained using the Heisenberg equation
\begin{equation}
\label{EQ1} 
\dot{\hat{\rho}}(t) = -\frac{i}{\hbar}\big[\hat{H}_I(t), \hat{\rho}(t)\big].
\end{equation}
This leads to the time evolution of the atomic subsystem for which the (reduced) density operator is given by taking partial trace with respect to filed Hilbert space, $\hat{\rho}_{A}=tr_{F}\hat{\rho}$. Formal solution of Eq.~(\ref{EQ1}) yields the 
integro-differential equation for atomic density operator as
\begin{eqnarray}\label{EQ2} 
&&\hspace*{-.2in}\dot{\hat{\rho}}_A(t)=-\frac{i}{\hbar}\trf\big[\hat{H}_I(t) ,\hat{\rho}(t_i)\big]\nonumber\\
&& \hspace*{.3in}- \frac{1}{\hbar^{2}} \int_{t_i}^{t} dt^\prime \trf \Big[\hat{H}_I(t),\big[\hat{H}_I(t^\prime) , \hat{\rho}(t^\prime)\big]\Big].
\end{eqnarray}
In the absence of atom-field interaction, the density operator $\hat{\rho}(t)$ reduces to the direct product of $\hat{\rho}_{A}(t)$ and $\hat{\rho}_{F}(t)$. Hence, the presence of interaction Hamiltonian induces a deviation that may be taken into account by adding an extra term $\hat{\rho}_{I}$,
\begin{equation}\label{S} 
\hat{\rho}(t)=\hat{\rho}_{A}(t)\hat{\rho}_{F}(t)+ \hat{\rho}_{I}(t).
\end{equation}
Obviously, $\trf[\hat{\rho}_{I}(t)]=0$. Moreover, choosing $t_i$ as the moment the atom-field interaction begins, $[\rho_I(t_i)=0]$, 
\begin{eqnarray}
\label{EQ3}
&&\hspace{-0.3in}\dot{\hat{\rho}}_{A}(t)=
-\frac{i}{\hbar}\trf\big[\hat{H}_I(t) , \hat{\rho}_{A}(t_i)\hat{\rho}_{F}(t_i)\big]\nonumber\\
&&- 
\frac{1}{\hbar^{2}} \int_{t_i}^{t} dt^\prime \trf \Big[\hat{H}_I(t),
\big[\hat{H}_I(t^\prime) , \hat{\rho}_{A}(t^\prime)\hat{\rho}_{F}(t')\big]\Big],
\end{eqnarray}
where we have ignored higher orders of $H_I$ assuming a weak atom-filed coupling, by dropping $\hat{\rho}_I(t')$ in the commutator within the integrand. It may be worth noting that the thermal field has been treated as a reservoir; it is  made up of oscillators of whole frequency range [Recall Eq.~(\ref{FH})]. Hence, $\hat{\rho}_{F}(t')$ in the integrand can be replaced by its initial value $\hat{\rho}_{F}(t_i) \equiv \hat{\rho}_{F}$, that for a thermal field of temperature $T$ reads
\begin{equation}
\label{FS} 
\hat{\rho}_F = \frac {\exp{\displaystyle \frac{-\hat {H}_F}{\kappa_B T}}}{\mathrm{tr} \exp \displaystyle{\frac{-\hat {H}_F}{\kappa_B T}}}~, 
\end{equation} 
where $\kappa_B $ is the Boltzamnn constant.
Further,
from Eq.~(\ref{IH}) together with Eqs.~(\ref{EF}) and (\ref{EF2}) it is easily seen that since the interaction Hamiltonian is linear in terms of photon annihilation and creation operators, the first term in the right hand side of Eq.~(\ref{EQ3}) vanishes as the commutator is off-diagonal with respect to the eigenstates of field Hamiltonian. The next step is to calculate the remaining commutators in Eq.~({\ref{EQ3}}). In order to facilitate the calculation, we rewrite the interaction Hamiltonian (\ref{IH}), using Eqs.~(\ref{di}) and (\ref{EF2}) and applying the rotating wave approximation as follows,
\begin{equation}
\label{H+-}
\hat{H}_I= \hat{H}^+_I + \hat{H}^-_I,
\end{equation}
where
\begin{equation}
\label{H+}
\hat{H}^+_I= - \int ^{\infty}_{0} d\omega  \, d_{\alpha} \hat{ E}_\alpha(\bm{r}_A , \omega) \hat{\sigma}^\dagger e^{i(\omega_A - \omega) t} ,\,\,\hat{H}^-_I = (\hat{H}^+_I)^\dagger\,. 
\end{equation}
Omitting traceless terms in the integrand in Eq.~(\ref{EQ3}) we end up with
\begin{eqnarray}
\label{EQ4}
&&\hspace*{-.2in}\dot{\hat{\rho}}_{A}(t)=\frac{1}{\hbar^{2}} \int_{t_i}^{t} dt^\prime \big[\hat X(t , t^\prime ) - \hat Y(t , t^\prime ) \nonumber\\
&&\hspace*{1in}-\hat Z(t , t^\prime )+ \hat W(t , t^\prime )\big]+ \hc ,
\end{eqnarray}
where
\begin{eqnarray}
\label{XYZW1}
&&\hat{X}(t , t^\prime) = \trf\big[\hat{H}^+_I(t)\hat{\rho}_{F}
\hat{\rho}_{A}(t')\hat{H}^-_I(t^\prime)\big],\nonumber\\
&&
\hat{Y}(t , t' )= \trf\big[\hat{H}^+_I(t) 
\hat{H}^-_I(t^\prime)\hat{\rho}_{F}\hat{\rho}_{A}(t^\prime)\big],\nonumber\\
&&\hat Z(t , t^\prime )= \trf\big[ \hat{\rho}_{F}\hat{\rho}_{A}(t^\prime)\hat{H}^-_I(t^\prime)\hat{H}^+_I(t)\big], \nonumber\\
&&
\hat W(t , t^\prime )= \trf\big[\hat{H}^-_I(t)\hat{\rho}_{F}\hat{\rho}_{A}(t^\prime)\hat{H}^+_I(t^\prime)\big].
\end{eqnarray}
In order to calculate $\hat{X}(t,t')$, for example, we use first Eq.~(\ref{H+}) from which we find
\begin{eqnarray}
&&\hspace*{-.4in}\hat{H}^+_I(t)\hat{\rho}_{F}\hat{\rho}_{A}(t^\prime)\hat{H}^-_I(t^\prime) = \int_0^\infty d\omega\int_0^\infty d\omega' d_\alpha d_\beta\nonumber\\
&&\hspace*{-.2in}\times 
\hat{\sigma}^\dagger \hat{E}_\alpha(\bm{r}_A, \omega)\hat{\rho}_F 
 \hat{\rho}_A(t')\hat{\sigma}\hat{E}_\beta^\dagger(\bm{r}_A ,\omega')
 e^{i(\Delta t - \Delta' t')}
\end{eqnarray}
with $ \Delta = \omega_A - \omega $ and $ \Delta' = \omega_A - \omega' $. Taking partial trace with respect to filed subsystem together with applying Eqs.~(\ref{IH})--(\ref{EF2}) leads to
\begin{eqnarray}
\label{EqA2}
&&
\hat{X}(t,t')= 
\int_0^\infty d\omega\int_0^\infty d\omega' d_\alpha d_\beta 
e^{i(\Delta t - \Delta' t')}\nonumber\\
&&\times\trf\big[
\hat{E}_\alpha(\bm{r}_A, \omega)\hat{\rho}_F 
 \hat{E}_\beta^\dagger(\bm{r}_A ,\omega')\big] \hat{\sigma}^\dagger \hat{\rho}_A(t')\hat{\sigma}
 \nonumber\\
&&
= 
\int_0^\infty d\omega\int_0^\infty   d\omega' \sum_\lambda\sum_{\lambda'}
\int d^3r \int d^3 r'e^{i(\Delta t - \Delta' t')}\nonumber\\
&&\times d_\alpha  d_\beta G_{\lambda \alpha i}(\bm{r}_A,\bm{r},\omega)
G^{\ast}_{\lambda' \beta j}(\bm{r}_A,\bm{r}',\omega') 
 \hat{\sigma}^\dagger \hat{\rho}_A(t')\hat{\sigma}\nonumber\\
&&\times
\trf \left[\hat{f}^\dagger_{\lambda' j}(\bm{r}', \omega')
\hat{f}_{\lambda i}(\bm{r},\omega)\hat{\rho}_F \right].
\end{eqnarray}
Using the fact that $\hat\rho_F$ is diagonal in the Hilbert space 
spanned by photonic eigenstates, it is inferred that the partial trace 
in the last line of Eq.~(\ref{EqA2}) vanishes except for the photon creation and annihilation 
operators with identical peer to peer labels. This results in the replacement
\begin{equation}
\label{EqA3}
\hat{f}^\dagger_{\lambda' j}(\bm{r}', \omega')
\hat{f}_{\lambda i}(\bm{r},\omega)
\to 
 \hat{n}(\omega) \delta_{\lambda \lambda^\prime}
\delta_{ij}  \delta(\bm{r} - \bm{r}^\prime )\delta(\omega- \omega^\prime)
\end{equation}
with $\hat{n}(\omega)$ being the operator of photon number (compare
with  $\hat{a}^\dagger_{\bm{k}} \hat{a}_{\bm{k}} = \hat{n}_{\bm{k}}$ for discrete field modes), which in turn leads to
\begin{eqnarray}
\label{EqA4}
&&\trf\left[ \hat{f}_{\lambda' j }^{\dagger} (\bm{r} , \omega )\hat{f}_{\lambda i }(\bm{r}^\prime , \omega^\prime ) \hat{\rho}_F \right] \nonumber\\
&&=\bar{n}(\omega) \delta_{\lambda \lambda^\prime}
\delta_{ij}  \delta(\bm{r} - \bm{r}^\prime )\delta(\omega- \omega^\prime).
\end{eqnarray}
Substitution of the trace from Eq.~(\ref{EqA4}) into Eq.~(\ref{EqA2})
we obtain
\begin{eqnarray}
\label{EqA5}
&&\hspace{-.4in}
\hat{X}(t,t')=
\int_0^\infty d\omega \bar{n}(\omega)
\hat{\sigma}^\dagger \hat{\rho}_A(t')\hat{\sigma}
 e^{i\Delta(t - t')}\nonumber\\
&&\hspace{-.3in}\times 
\sum_\lambda
\int d^3r \,\bm{d} \cdot \bm{G}_{\lambda }(\bm{r}_A,\bm{r},\omega)\cdot
\bm{G}^{\ast \top}_{\lambda}(\bm{r}_A,\bm{r},\omega)\cdot
\bm{d}\,.
\end{eqnarray}
From Eq.~(\ref{EqA5}) by performing the position integral as well as the summation over $\lambda$ using the integral relation
(\ref{IE}), we find
 \begin{eqnarray}
 \label{X2}
&&\hat X(t , t^\prime ) = \frac{\hbar \mu_0}{\pi} \int^{\infty}_{0}d\omega  \bar{n}(\omega)\omega^2 \bm{d} \cdot \mathrm{Im} \bm{G}(\bm{r}_A , \bm{r}_A , \omega) \cdot \bm{d} \,\nonumber\\
&&\hspace*{1in}\times\hat{\sigma}^\dagger \hat{\rho}_{A}(t^\prime) \hat{\sigma} e^{i(\omega_A - \omega)(t-t^\prime) },
\end{eqnarray}
where $\bar{n}(\omega)$ $\! =$ $\! 1/[e^{\hbar \omega/(k_B T)}$ $\!-$ $\!1]$ is the mean photon number of the thermal field.
Calculating $\hat Y$, $\hat Z$, and $\hat W$ in a similar manner and substitution of the resulting expressions into Eq.~(\ref{EQ4}) leads to
\begin{eqnarray}
\label{X3}
&&\hspace*{-.2in} \dot{\hat{\rho}}_{A}(t)=\frac{\mu_0}{ \hbar \pi} \int_{t_i}^{t} dt^\prime \int^{\infty}_{0}d\omega \omega^2 \bm{d} \cdot \mathrm{Im} \bm{G}(\bm{r}_A , \bm{r}_A , \omega)
\cdot\bm{d}  \nonumber \\
&& \hspace*{-.1in} \times \Big\{
\big[\bar{n}(\omega)+1\big] \left[\hat{\sigma} \hat{\rho}_{A}(t^\prime)\hat{\sigma}^\dagger - \hat{\sigma}^\dagger  \hat{\sigma}\hat{\rho}_{A}(t^\prime) \right]\nonumber\\
&&
\hspace*{.2in}+
\bar{n}(\omega)  \left[\hat{\sigma}^\dagger \hat{\rho}_{A}(t^\prime)\hat{\sigma} - \hat{\rho}_{A}(t^\prime)\hat{\sigma} \hat{\sigma}^\dagger\right] \Big\}\, e^{i(\omega_A - \omega)(t-t^\prime) }\nonumber\\
&&\hspace*{2.2in} + \textrm{H.c.}\,.
\end{eqnarray} 

In the weak-coupling regime of atom-field interaction a notable change in atomic subsystem requires appreciable period of time.
Thus, we may set the lower limit of the time integral in Eq.~(\ref{X3}) to $-\infty$ with negligible error. On the other hand, it is seen that the time evolution of the atomic density operator at time $t$, depends on its value in all previous times, $\hat{\rho}(t')$, which is slowly varying compared to the exponential factor
$e^{i(\omega_A - \omega)(t-t^\prime) }$ existing in the integrand. Hence, we may apply Markov approximation
by replacing $\hat{\rho}(t')$ by $\hat{\rho}(t)$ in the right hand side. So far,
\begin{eqnarray}
\label{EQ51}
&& \hspace*{-7ex}\dot{\hat{\rho}}_{A}(t)=\frac{\mu_0}{ \hbar \pi}  \int^{\infty}_{0}d\omega 
\int_{-\infty}^{t} dt^\prime e^{i(\omega_A - \omega)(t-t^\prime) }\nonumber \\
&&\hspace*{-7ex}\times\bigg(\omega^2 \bm{d} \cdot \mathrm{Im} \bm{G}(\bm{r}_A , \bm{r}_A , \omega)
\cdot\bm{d}
\nonumber\\
&&\hspace*{-3ex}\times\Big\{\bar{n}(\omega) \big[\hat{\sigma}^\dagger \hat{\rho}_{A}(t)\hat{\sigma} - \hat{\rho}_{A}(t)\hat{\sigma} \hat{\sigma}^\dagger\big]
  \nonumber\\
  && \hspace*{-3ex}+ \big[\bar{n}(\omega)+1\big] \big[\hat{\sigma} \hat{\rho}_{A}(t)\hat{\sigma}^\dagger - \hat{\sigma}^\dagger  \hat{\sigma}\hat{\rho}_{A}(t) \big]\Big\} \bigg)+ \hc\,.
\end{eqnarray} 
The remaining time integral is divergent. The divergence can be removed by altering
the atomic transition frequency with an infinitesimal positive imaginary part that corresponds to giving a spectral width to the excited state. In summary,
\begin{eqnarray}
\label{Eq40}
&&\int_{t_i}^t dt^\prime e^{i(\omega_A -\omega)(t - t^\prime)} \to \lim_{\eta \to 0^+} \int_{-\infty}^t dt^\prime e^{i(\omega_A -\omega +i \eta)(t - t^\prime)}
\nonumber\\
&& \hspace*{1in}= \pi\delta(\omega_A -\omega)+i \textrm{P}\frac{1}{\omega_A -\omega}
\end{eqnarray}
with P standing for principal value.
Making use of Eq.~(\ref{Eq40})  into Eq.~(\ref{EQ51}) leads to Eq.~(\ref{EQ6}) together with Eq.~(\ref{gamma0}). 
Note that the pure imaginary principal value integral resulting from the second term in the right hand side of Eq.~(\ref{Eq40}) is cancelled as its Hermitian conjugate is present in Eq.~(\ref{EQ51}).  

At this stage, performing the above process, we are left with the differential equation governing the atomic density operator as
\begin{eqnarray}
\label{EQ6}
&&\hspace*{-.5in}\dot{\hat{\rho}}_{A}(t)=\frac{\Gamma_0}{2} 
\left\{ (\bar{n}+1) \left[\hat{\sigma} \hat{\rho}_{A}(t)\sigma^\dagger - \hat{\sigma}^\dagger  \hat{\sigma}\hat{\rho}_{A}(t) \right]\right.\nonumber\\
&&\left. + \bar{n} \left[\hat{\sigma}^\dagger \hat{\rho}_{A}(t)\hat{\sigma} - \hat{\rho}_{A}(t)\hat{\sigma} \hat{\sigma}^\dagger\right]\right\} + \hc
\end{eqnarray} 
$ [\bar{n}\equiv \bar{n}(\omega_A)]$ with 
\begin{equation}\label{gamma0}
\Gamma_0=\frac{2 \omega^2_A}{\hbar \epsilon_0 c^2}\bm{d} \cdot \mathrm{Im} \bm{G}(\bm{r}_A , \bm{r}_A , \omega_A) \cdot \bm{d}
\end{equation}
being the madium-assisted decay rate of an excited atom at zero temperature
\cite{dung2003electromagnetic}. Its counterpart for non-zero temperatures can be obtained making use of Eq.~(\ref{EQ6}).
To this end, denoting by $p_e$ and $p_g$ the probability of atom being, respectively, in excited state and ground state, we arrive in a set of coupled differential equations,
 \begin{equation}
 \label{pdot1}
 \dot{p}_e (t)= \langle e| \dot{\hat{\rho}}_A(t) | e \rangle =
  \Gamma_0 [ \bar{n}p_g (t)-(\bar{n}+1)p_e (t)],  
 \end{equation}
 \begin{equation}
 \label{pdot2}
   \dot{p}_g(t)=\langle g| \dot{\hat{\rho}}_A(t) | g \rangle = -  \dot{p}_e (t).
 \end{equation}

We conclude this section by considering the case of atom being initially in the excited state for which,
 solving Eqs.~(\ref{pdot1}) and (\ref{pdot2}) leads to
 \begin{equation}
 p_e(t) = \frac{1}{\mathbb{Z}}
 \left\{e^{-E_g/(k_B T)}e^{-\Gamma_T t} + 
e^{-E_e/(k_B T)} \right\},
 \end{equation} 
where $\mathbb{Z} =  e^{-E_g/(k_B T)}+e^{-E_e/(k_B T)}$ 
is the partition function
and \begin{equation}
\label{Gamma}
\Gamma_T = [2\bar{n}(\omega_A)+1]\Gamma_0
\end{equation}
is the medium-assisted decay rate of an excited atom at temperature $T$. 
As expected it reduces to the well-known zero-temperature result for $T$ $\!=$  $\!0$ $(\bar{n}$ $\!=$ $\!0)$. Moreover, the effect of non-zero temperature is found to be equal to the decay rate in zero temperature multiplied by two times of the photons number present at the frequency of atomic transition.

\section{Summary and conclusion}
\label{summ}
In this paper the time evolution of a two-level atom interacting with an electromagnetic field in thermal state
has been explored by the density operator approach, through which the decay rate of an excited atom is formulated.  The result is a generalization of previous findings, by regarding both the presence of arbitrarily arranged magneto-electric media and the non-zero temperature of the field. It turned out that the non-zero temperature contributes to the decay rate via an additive term that is equal to the zero-temperature decay rate 
multiplied by two times of mean photon number present in the thermal filed at the frequency of atomic transition.

It is worth noting that in the calculations we have considered weak-coupling regime of 
atom-filed interaction. The cases where the atom is interacting near-resonantly with an 
electromagnetic field mediated by a resonator-like surrounding, throws doubt on the validity of Markov 
approximation. The case of a two-level atom strongly coupled to a resonant environment has been developed in terms of exact master equation describing a non-Markovian decay in Ref.~\cite{garraway1997decay}, where the photon density of state is modelled as Lorentzian narrow-band 
non-monochromatic modes. To account for magneto-electric and geometric properties of surrounding media a possible solution is to include
properly the Green's function to the modelled electromagnetic field spectrum,
for example, the way performed in the calculation of
the Casimir-Polder force in strong atom-field 
interaction in Ref.~\cite{buhmann2008strong}. 


\end{document}